\documentclass[conference]{IEEEtran}
\IEEEoverridecommandlockouts
\usepackage{cite}
\usepackage{amsmath,amssymb,amsfonts}
\usepackage{algorithmic}
\usepackage{graphicx}
\usepackage{textcomp}
\usepackage{xcolor}
\usepackage{lipsum}
\usepackage{url}
\usepackage{comment}
\usepackage{todonotes}
\usepackage{tikz}
\usepackage[utf8]{inputenc}
\usepackage{soul}
\usetikzlibrary{arrows}

\definecolor{gray}{rgb}{0.7,0.7,0.7}
\definecolor{orange}{rgb}{1.0,0.5,0.0}
\definecolor{new}{rgb}{1.0,0.2,0.0}
\definecolor{old}{rgb}{0.0,0.0,0.8}
\newcommand{\dummy}[1]{{\textcolor{gray}{#1}}}

\newcommand{\rv}[1]{{\textcolor{black}{#1}}}
\newcommand{\key}[1]{\textbf{#1}}

\newcommand{\rone}{\textbf{(N1.1)}}
\newcommand{\rtwo}{\textbf{(N1.2)}}
\newcommand{\rthree}{\textbf{(N1.3)}}
\newcommand{\rfour}{\textbf{(N1.4)}}

\newcommand{\none}{\textbf{(N2.1)}}
\newcommand{\ntwo}{\textbf{(N2.2)}}
\newcommand{\nthree}{\textbf{(N2.3)}}
\newcommand{\nfour}{\textbf{(N2.4)}}

\newcommand{\oone}{\textbf{(N3.1)}}
\newcommand{\otwo}{\textbf{(N3.2)}}
\newcommand{\othree}{\textbf{(N3.3)}}
\newcommand{\ofour}{\textbf{(N3.4)}}

\newcommand{\lrone}{\textbf{(L1.1)}}
\newcommand{\lrtwo}{\textbf{(L1.2)}}
\newcommand{\lrthree}{\textbf{(L1.3)}}
\newcommand{\lrfour}{\textbf{(L1.4)}}

\newcommand{\lnone}{\textbf{(L2.1)}}
\newcommand{\lntwo}{\textbf{(L2.2)}}
\newcommand{\lnthree}{\textbf{(L2.3)}}
\newcommand{\lnfour}{\textbf{(L2.4)}}

\newcommand{\loone}{\textbf{(L3.1)}}
\newcommand{\lotwo}{\textbf{(L3.2)}}
\newcommand{\lothree}{\textbf{(L3.3)}}
\newcommand{\lofour}{\textbf{(L3.4)}}

\newcommand{\drawref}[6]{\fill[#5] (#1,#2) ellipse (0.05cm and 0.05cm);\draw (#1+#3,#2+#4) node {\footnotesize \textcolor{#5}{#6}};}

\def\BibTeX{{\rm B\kern-.05em{\sc i\kern-.025em b}\kern-.08em
    T\kern-.1667em\lower.7ex\hbox{E}\kern-.125emX}}

\begin{document}

\title{A Survey on Computing Schematic Network Maps: \\ The Challenge to Interactivity}

\author{%
\IEEEauthorblockN{Hsiang-Yun Wu}
\IEEEauthorblockA{%
\textit{TU Wien, Austria} \\
hsiang.yun.wu@acm.org}
\and
\IEEEauthorblockN{Benjamin Niedermann}
\IEEEauthorblockA{%
\textit{University of Bonn, Germany}\\
niedermann@igg.uni-bonn.de}
\and
\IEEEauthorblockN{Shigeo Takahashi}
\IEEEauthorblockA{%
\textit{University of Aizu, Japan}\\
takahashis@acm.org}
\and
\IEEEauthorblockN{Martin N{\"o}llenburg}
\IEEEauthorblockA{%
\textit{TU Wien, Austria} \\
noellenburg@ac.tuwien.ac.at}
}

\maketitle

\begin{abstract}
Schematic maps are in daily use to show the connectivity of subway systems and to facilitate travellers to plan their journeys effectively. 
This study surveys up-to-date algorithmic approaches in order to give an overview of the state of the art in schematic network mapping. 
\rv{The study investigates the hypothesis that the choice of algorithmic approach is often guided by the requirements of the mapping application. For example, an algorithm that computes globally optimal solutions for schematic maps is capable of producing results for printing, while it is not suitable for computing instant layouts due to its long running time.
Our analysis and discussion, therefore, focus on the computational complexity of the problem formulation and the running times of the schematic map algorithms, 
including algorithmic network layout techniques and station labeling techniques. 
The correlation between problem complexity and running time is then visually depicted using scatter plot diagrams.}
Moreover, since metro maps are common metaphors for data visualization, we also investigate online tools and application domains using metro map representations for analytics purposes, and finally summarize the potential future opportunities for schematic maps.
\end{abstract}


\section{Introduction} \label{sec:intro}


A metro map is a schematic visual representation of an underlying transit network that depicts the connectivity between metro stations and lines of a public transportation system~\cite{o-mmw-03}.
With well-designed metro maps, travellers can effectively identify their locations and find their way or perform routing planning on a complex transportation system in a big city, such as London, Paris, or Tokyo.
These maps are especially helpful since travellers often look for a quick solution for the shortest or cheapest path from station $A$ to $B$, how to transfer from $A$ to $B$, and how many stations left to $B$~\cite{r-umu-12}.
To support these tasks, Henry Beck introduced the so-called Tube Map of London Underground, and has proposed several drawing criteria to achieve this~\cite{g-mbum-94}, \rv{and many extended versions have been evaluated~\cite{Roberts:2017:HCS}\cite{Roberts:2013:HCS}.}

Nonetheless, the need for automatic drawing algorithms is still increasing due to the high cost and limited adaptability of hand-drawn maps.
This has been considered as a difficult problem because several subproblems, including layout schematization, line crossing minimisation, and map label placement have been studied and proved as complex problems~\cite{noellenburg:2014:schematicmapping}.
Two state-of-the-art reports from 2007 and 2014 have surveyed similar approaches before~\cite{w-dsms-07}\cite{noellenburg:2014:schematicmapping}. Hence we focus on relatively new approaches from the last 5 years after N{\"o}llenburg's report~\cite{noellenburg:2014:schematicmapping} and discuss the correlation in terms of computational complexity and interactivity, as well as the potential techniques that can be used in different research domains.

\subsection{Problem Definition}

The initial idea for a metro map lies on simplifying the layout geometry to facilitate users' comprehensive understanding of connectivity between metro stations.
This allows us to formulate the drawing problem as network visualization problems, which are often studied to untangle visual clutter of the layout to improve its readability~\cite{w-gdc-13}.

\subsubsection{Metro Map Problem}

Let us formulate the metro map problem by introducing an undirected graph $G=(V, E)$ embedded in a plane $\mathbb{R}^2$.
Each vertex $v \in V$ represents a metro station and an edge $e=(v_i,v_j) \in E$ indicates the physical connection between two metro stations. In addition, a line $l \in L$ is a metro line containing a set of metro stations and edges that are defined from the corresponding metro system. Note that $l$ is a set cover, which implies each edge $e$ belongs to at least one $l \in L$. We call $MG=(G,L)$ here a \emph{Metro Graph} as previously defined by N{\"o}llenburg~\cite{noellenburg:2014:schematicmapping}. 
The input of the proposed problem is the connectivity of a Metro Graph $MG$ together with its geographically accurate embedding, 
and the solution aims to find a schematic layout of $MG$ that maximally satisfies several user-defined aesthetic drawing criteria.

\subsection{Our Taxonomy Design}

\rv{In this study, we investigate the trade-off between computation times and layout quality as well as application requirements.
This 
is motivated by our observation that a user who is trying to create a map would accept longer computation times for an exact solution if the generated map is expected to be printed,
while the user will not be patient if the application is expected to be an interactive application.
To investigate our assumption systematically, 
we analyze several factors in this survey (see Tables~\ref{tab:score_net} and~\ref{tab:score_label}) that influence the computational complexity and optimality requirements of a problem together with the corresponding running time of the algorithmic approach (Table~\ref{tab:score_label}).
Our two primary topics cover algorithmic network layout techniques and station labeling techniques. 
The correlation between problem complexity and time complexity is then given using a scatter plot diagram visually.}

\rv{This is done by categorizing publications along two primary coordinates, problem complexity and running time.}
These include (1) the range from local to global optimality in terms of the incorporated aesthetic criteria and objective functions, and (2) the range of suitability from static to interactive visualizations based on the computation speed.
The values on the first coordinate are computed using the scoring Tables~\ref{tab:score_net} and~\ref{tab:score_label}. 
\rv{For simplicity, we assume that the degree of interactivity, running time in other words, of a schematic map algorithm is potentially linearly-correlated to the degree of criteria selected by the proposed approach.}
In other words, we expect the developed techniques would find a good balance between the efficiency of the algorithms and the quality of the schematic maps generated by those approaches.
For better instructing readers along this assumption, we will focus our discussion on these two aspects in the following sections.


\subsection{Tasks and Design Rules}

It is known that drawing criteria are often designed based on users' effectiveness to accomplish tasks on a map.
These tasks are similar to tasks on graphs, such as \emph{Topology-Based Tasks}, \emph{Attribute-Based Tasks}, and \emph{Browsing Tasks} since finding shortest or cheapest paths, identifying a station of a line, and navigating along a specific route are mostly performed by the travellers.
We revisit N{\"o}llenburg's list~\cite{noellenburg:2014:schematicmapping} of design principles and investigate which of these serve as dominant constraints, for generating a globally optimal map, in the coming sections.
In this survey, we primarily focus on two directions in this field, including network layout techniques {\textbf{(N)}} and labeling techniques {\textbf{(L)}} as summarized in Table~\ref{tab:score_net} and Table~\ref{tab:score_label}, respectively.
%
%
%
%
%
%
%
%
%
Note that the selected criteria are ordered in the sense that the criterion with higher scores influences global structures of the layout, while the one with lower scores affects the layout in a local fashion.
This scoring scheme will then be used as an indicator for guiding readers to navigate the diagrams created in the coming sections.
Based on the aforementioned scoring scheme, we derive our novel taxonomy of metro map techniques in this survey paper.

The remainder of this paper is structured as follows. In Section~\ref{sec:network}, we summarize relevant schematic network layout algorithms, and in Section~\ref{sec:feature}, research approaches integrating map features, mainly on text and image labels.
In Section~\ref{sec:app}, we then introduce several tools that are accessible online and demonstrate the usability of the collected techniques in different research domains.
Finally in Section~\ref{sec:conclude}, we conclude this paper and list several future directions to this topic.


%
%

\section{Algorithmic Network Layout Techniques} \label{sec:network}


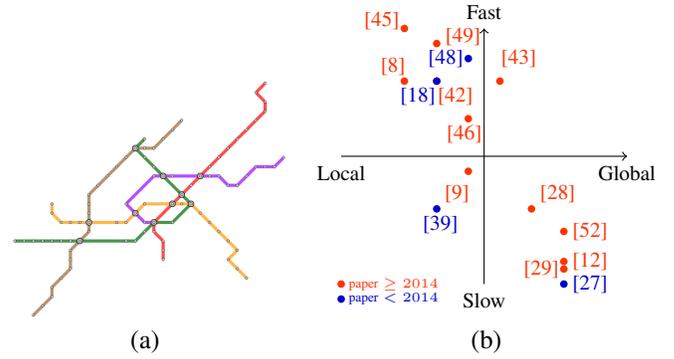
\begin{figure}[htb]
\centering{
 \setlength{\tabcolsep}{0pt}
 \begin{tabular}{cc}
    \includegraphics[width=0.49\linewidth]{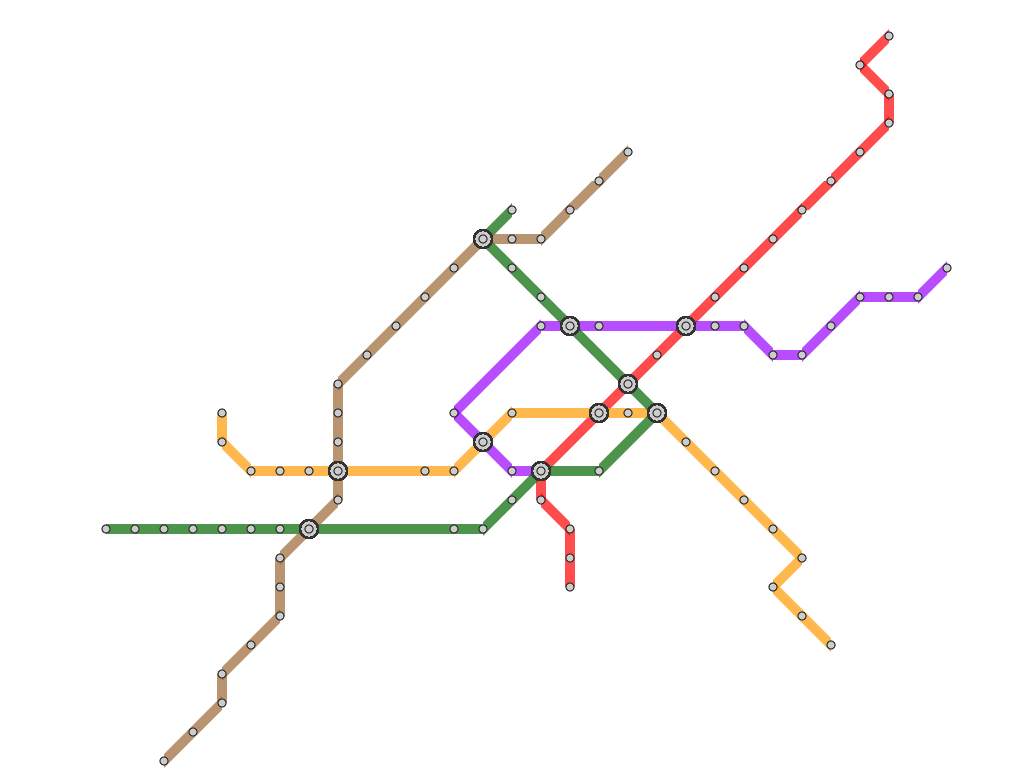} &
    \begin{tikzpicture}[
    scale=1,
    axis/.style={ ->, >=stealth'},
    dashed line/.style={dashed, thin},
    every node/.style={color=black}
    ]
    
    \draw[->] (-1.9,0) node(xline)[below] {\footnotesize Local} -- (1.9,0) node(xline)[below] {\footnotesize Global};
    \draw[->] (0,-1.7) node(yline)[below] {\footnotesize Slow}-- (0,1.7) node(yline)[above] {\footnotesize Fast};
    \drawref{-0.63}{1.0}{0.2}{-0.2}{new}{\cite{Ti2014}} 
    \drawref{0.63}{-0.7}{0.3}{0.2}{new}{\cite{Oke2015}} 
    \drawref{-0.63}{1.5}{0.3}{0.1}{new}{\cite{WangPeng2016}} 
    \drawref{1.06}{-1.0}{0.3}{0.0}{new}{\cite{Wu2015}} 
    \drawref{0.21}{1.0}{0.2}{0.3}{new}{\cite{Ti2015}} 
    \drawref{1.06}{-1.5}{-0.35}{0.0}{new}{\cite{Onda2018}}
    \drawref{-0.21}{0.5}{-0.1}{-0.2}{new}{\cite{Dijk2014}} 
    \drawref{-1.06}{1.0}{-0.2}{0.2}{new}{\cite{Chivers2014}} 
    \drawref{-0.21}{-0.2}{-0.2}{-0.3}{new}{\cite{Chivers2015}} 
    \drawref{-1.06}{1.7}{-0.35}{0.1}{new}{\cite{Dijk2018}} 
    \drawref{-0.63}{1.0}{-0.3}{-0.20}{old}{\cite{hmn-avmm-06}} 
    \drawref{-0.63}{-0.7}{0.0}{-0.20}{old}{\cite{srmw-amlumo-11}}
    \drawref{1.06}{-1.7}{0.3}{0.0}{old}{\cite{nw-dlhqm-11}}
    \drawref{1.06}{-1.4}{0.3}{0.0}{new}{\cite{Fink:2014:SMW}}
    \drawref{-0.21}{1.3}{-0.35}{0.0}{old}{\cite{wc-fmm-11}}
    \drawref{-1.9}{-1.7}{0.7}{0.0}{new}{\tiny{paper $\ge 2014$}}
    \drawref{-1.9}{-1.9}{0.7}{0.0}{old}{\tiny{paper $<2014$}}
    \end{tikzpicture} 
    \\
  (a)&(b)\\
 \end{tabular}
}
\caption{Network map visualization, including an example of (1) Vienna metro map~\cite{nw-dlhqm-11}, and (b) layout techniques with respect to their globality and potential for interactiveness.}
\label{fig:tax1}
\end{figure}


%
\begin{table}[thb]
\vspace*{-3mm}
\caption{The point system for evaluating the criterion effectiveness.}
\centering{
\begin{tabular}{|c|c|l|} \hline
ID & S & Description (x-coordinate in Figs.~\ref{fig:tax1}-~\ref{fig:tax3}) \\ \hline
(\textbf{N1})&  & \textbf{Combinatorial property}\\ 
\rone{}     & 4     & Overall combinatorial optimization. \\  
\rtwo{}     & 3     & Combinatorial criteria for sets of vertices or edges. \\  
\rthree{}   & 2     & Combinatorial criteria for pairs of vertices or edges. \\ 
\rfour{}    & 1     & Combinatorial criteria for single vertices or edges. \\ \hline \hline 
%
(\textbf{N2}) & & \textbf{Geometry property}\\
\none{}     & 4     & Uniform geometric optimization. \\  
\ntwo{}     & 3     & Geometric criteria for sets of vertices or edges. \\  
\nthree{}   & 2     & Geometric criteria for pairs of vertices or edges. \\  
\nfour{}    & 1     & Geometric criteria for single vertices or edges. \\ \hline \hline 
%
%
(\textbf{N3}) & & \textbf{Approach optimality}\\
\oone{}     & 4     & Global optimality and exact global optimization. \\  
\otwo{}     & 3     & Global optimality, but local optimization. \\  
\othree{}   & 2     & Local optimality, but global optimality for sub-problems. \\  
\ofour{}    & 1     & Local optimality, and local optimization. \\ \hline 
\end{tabular}
}
\label{tab:score_net}
\end{table}
\begin{table}[thb]
\caption{The point system for evaluating the time effectiveness.}
\centering{
\begin{tabular}{|c|c|l|} \hline
ID & S & Time complexity (y-coordinate in Figs.~\ref{fig:tax1}-~\ref{fig:tax3}) \\ \hline
\key{T1}     & 4     & The layout is computed less than a second. \\  
\key{T2}     & 3     & The layout is computed in a few seconds. \\ 
\key{T3}     & 2     & The layout is computed within a few minutes . \\ 
\key{T4}     & 1     & The layout is computed in several hours or more. \\ \hline 
\end{tabular}
}
\vspace*{-3mm}
\label{tab:score_time}
\end{table}
In this section we classify the different network layout algorithms proposed in the literature over the past 15 years along three criteria that determine the degree of globality of the algorithmic techniques. 
These criteria comprise combinatorial properties \key{(N1)} of the underlying graph structure (e.g., topology preservation), geometric properties \key{(N2)} of the input network and the schematic layout (e.g., directional deviations and line straightness), and the degree of optimality \key{(N3)} of the computed solutions, i.e., whether an global optimum is computed or just a local optimum.
Table~\ref{tab:score_net} lists the three criteria, which are ranked by scores ranging from 1 (high locality) to 4 (high globality). 
Additionally, we assess each technique by the required computational resources and resulting degree of suitability for interactive applications (summarized in Table~\ref{tab:score_time}).

Each of the papers discussed in the following paragraphs thus receives two scores that we use as coordinates in the two-dimensional scatter plot of Figure~\ref{fig:tax1}.

We start with four classic papers that serve as representatives of the main algorithmic techniques that were already discussed in the 2014 survey of Nöllenburg~\cite{noellenburg:2014:schematicmapping}. \rv{Since our focus is on new results from 2014 onward, we refer to the 2014 survey for a comprehensive discussion of the earlier literature.}
The force-based algorithm of Hong et al.~\cite{hmn-avmm-06} considers combinatorial and geometric parameters of individual and pairs of edges and vertices in their objective function, e.g., the slope of edges, or the preservation of the local network topology at each vertex. The objective function considers the sum of several forces acting upon each vertex to push it toward a state of locally minimal energy. Layouts are computed within a few seconds.

Stott et al.~\cite{srmw-amlumo-11}  define an explicit global objective function for metro maps, which takes smaller values for maps with higher layout quality. This quality function measures, e.g., angles at each vertex or between neighboring edges, local edge length differences, or the octilinearity of edges, which corresponds to a score of 2 in these criteria. The optimization itself is by a hill climbing algorithm that moves vertices to the best position in a local neighborhood, but evaluates the global map quality. The reported computation times range between a few minutes and a few hours.

The mixed-integer linear programming (MIP) approach of N{\"o}llenburg and Wolff~\cite{nw-dlhqm-11} defines combinatorial and geometric constraints among pairs of edges and vertices to guarantee, e.g., the correct network topology, octilinear edges, or a bounded amount of angular distortion, as well as the minimization of bends along lines and subpaths or the total network length. Their solution model, however, guarantees a globally optimal solution. To achieve this highest degree of optimality, one needs to invest computation times ranging from several minutes for simple network maps to several hours for more complex instances.

Finally, Wang and Chi~\cite{wc-fmm-11} define an energy function to represent the layout quality, with the goal of minimizing this energy function. As in the previous methods, this comprises aspects defined on pairs of vertices and edges such as edge lengths, angles, or octilinearity. Their optimization aims for global optimality, but using an iterative conjugate gradient method for least squares minimization to converge to a global minimum. Their algorithms typically run in less than a second.

Among the more recent papers, we mostly find approaches that improve and extend one of the classic techniques like MIP, local optimization heuristics, force-based iterative methods, or stroke-based incremental algorithms. 
Chivers and Rodgers~\cite{Chivers2014} present a hybrid force-based method that defines both a set of classic spring embedder forces~\cite{fr-gdfp-91} as well as a magnetic force field that models octilinearity~\cite{sm-gdmsm-95}. Their hybrid idea is to first let the spring forces find a well distributed layout and then let the magnetic forces become dominant to achieve octilinear edges. By the nature of their method they aim for locally optimal stable layouts, but the convergence is quite fast and typical metro maps can be computed in less than a second.
A second and very fast approach with similar properties is given by van Dijk and Lutz~\cite{Dijk2018}. They propose a method based on linearized least-squares minimization as an approximation of a non-linear model for computing linear cartograms, which are network drawings with prescribed edge lengths. This idea is then applied to metro map layout as one problem instance, where the length and slope of edges are soft constraints to be optimized. While several typical constraints of metro maps are not considered in their algorithm, it is an extremely fast technique with running times in the range of a few milliseconds.
Wang and Peng~\cite{WangPeng2016} provide another system using a global energy function whose minima correspond to locally optimal metro layouts in their model. The optimization technique is least-squares minimization. It comprises both octilinear and curvilinear layouts and is designed for interactive editing by a user who can modify the positions of a few \emph{handle} vertices. It is very fast and only requires a few milliseconds to compute medium-sized metro maps.

Three papers take the idea of decomposing the network into paths (also called \emph{strokes}), computing schematic representations of the strokes and then composing them into a single network again, see e.g.~\cite{ld-smagsnm-10}. Ti and Li~\cite{Ti2014} and Ti et al.~\cite{Ti2015} both propose an approach that first detects and enlarges areas in the geographically accurate representation of the metro network that have high vertex or edge density and then apply multifocal fisheye transformations to get a more uniform spatial distribution of the underlying network. In a second step, strokes are identified and locally schematized in an octilinear fashion. While no running times are reported in the papers, these methods are typically quite fast and run within at most a few seconds. On the other hand, the schematization only minimizes local distances to the input geometry and no explicit global objective function is optimized.
The algorithm by van Dijk et al.~\cite{Dijk2014} proposes schematic maps using strokes that are represented as circular arcs, and not as octilinear paths. They first simplify sequences of edges of the same metro line into longer paths, even across junctions, to reduce overall complexity of the layout and then find locally optimal representations of these paths/strokes by circular arcs guided by the Fréchet distance between paths and circular arcs.

Chivers and Rodgers~\cite{Chivers2015} extend the search space of the hill climbing local search technique~\cite{srmw-amlumo-11} by parameterizing grid spacing, local neighborhood size for node movements, and the number of iterations. The constraints and objectives are the same as those by Stott et al.~\cite{srmw-amlumo-11}. With their improvements they gain a speed-up factor between 5 and 8 compared to the earlier multi-criteria hill climbing technique~\cite{srmw-amlumo-11} and obtain the final (labeled) layouts within 5 to 60 minutes.

Finally, the MIP model has been improved and accelerated by several authors. Oke and Siddiqui~\cite{Oke2015} relax some of the integrality constraints of Nöllenburg and Wolff~\cite{nw-dlhqm-11} and drop one term in the objective function. Their model improves the running time by up to one order of magnitude to computation times of few minutes. Onda et al.~\cite{Onda2018} also modify the existing MIP model~\cite{nw-dlhqm-11}. They split the computation into two phases in order to speed up computation of large instances such as the Tokyo metro map. In the first phase they generate a rough layout that satisfies all hard constraints in an incremental face-by-face manner. The second phase optimizes the directions of short subpaths while keeping the directions of the remaining layout fixed. A layout of the Tokyo network was computed within 5 hours.
Certain topological structures can be highlighted by Wu et al., who allow for straightening a user-specified path~\cite{wtly-tmla-12}, or deforming a cyclic path into a circle~\cite{Wu2015} in order to emphasize meaningful structures. This has been achieved by introducing additional constraints to the standard MIP model, which increases the time for finding the corresponding solutions by a small amount.
\rv{Fink et al.~\cite{Fink:2014:SMW} use the MIP technique to model layout of concentric metro maps, where the network consists of polylines composed of radial or circular segments. The optimization considers among others bend minimization, minimization of distinct radial slopes, or uniformity of edge lengths. The reported running times are in the range of several minutes for small input maps like Vienna and Montreal.}








\section{Labeling Techniques} \label{sec:feature}



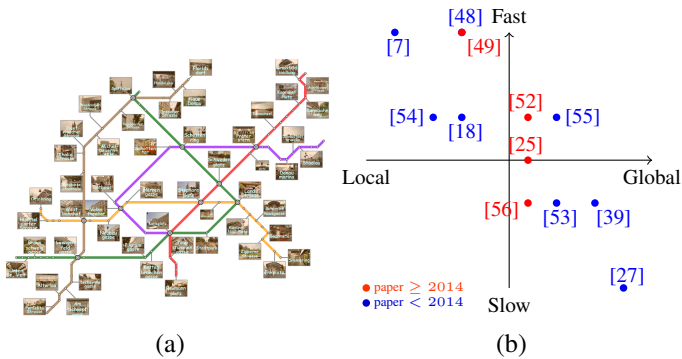
\begin{figure}[h]
\centering{
    \setlength{\tabcolsep}{0pt}
    \begin{tabular}{cc}
    \includegraphics[width=0.49\linewidth]{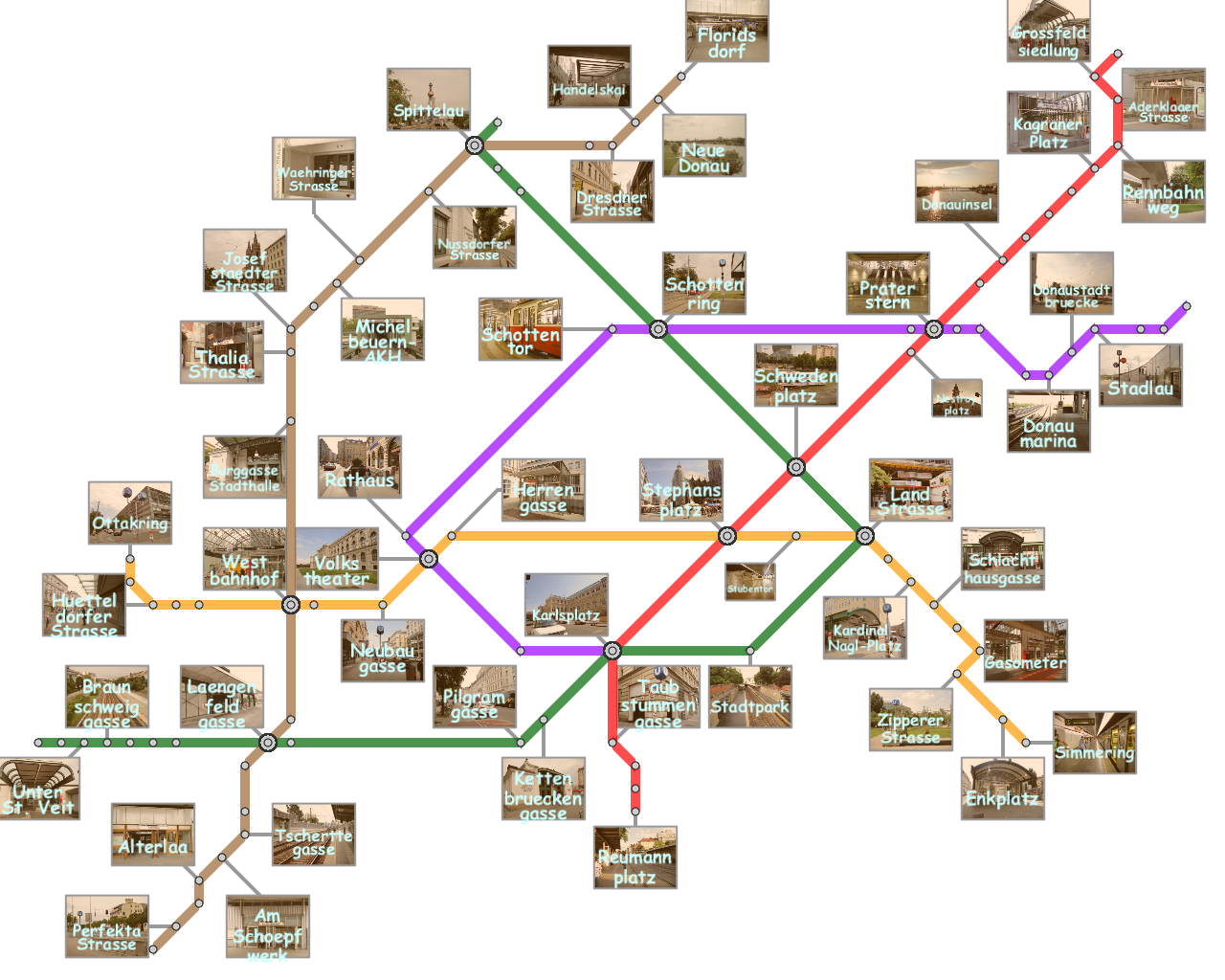} &
    \begin{tikzpicture}[
    scale=1,
    axis/.style={ ->, >=stealth'},
    dashed line/.style={dashed, thin},
    every node/.style={color=black}
    ]
    
    \draw[->] (-1.9,0) node(xline)[below] {\footnotesize Local} -- (1.9,0) node(xline)[below] {\footnotesize Global};
    \draw[->] (0,-1.7) node(yline)[below] {\footnotesize Slow}-- (0,1.7) node(yline)[above] {\footnotesize Fast};
    \drawref{0.25}{-0.57}{-0.4}{-0.1}{red}{\cite{yoshida2018}} 
    \drawref{0.25}{0}{-0.08}{0.2}{red}{\cite{Niedermann2018}} 
    \drawref{0.25}{0.57}{-0.08}{0.2}{red}{\cite{Wu2015}} 
    \drawref{0.63}{-0.57}{0}{-0.2}{blue}{\cite{wthaly-sedamm-13}} 
    \drawref{0.63}{0.57}{0.3}{0}{blue}{\cite{wtly-tmla-12}} 
    \drawref{-0.63}{1.7}{0}{0.20}{blue}{\cite{wc-fmm-11}} 
    \drawref{1.52}{-1.7}{0}{0.15}{blue}{\cite{nw-dlhqm-11}} 
    \drawref{-1.52}{1.7}{0}{-0.2}{blue}{\cite{cr-gidsmd-11}} 
    \drawref{1.14}{-0.57}{0.2}{-0.2}{blue}{\cite{srmw-amlumo-11}} 
    \drawref{-0.63}{0.57}{0.0}{-0.20}{blue}{\cite{hmn-avmm-06}} 
    \drawref{-1.01}{0.57}{-0.4}{0}{blue}{\cite{wtly-zapalmm-11}} 
    \drawref{-0.63}{1.7}{0.25}{-0.2}{red}{\cite{WangPeng2016}} 
    \drawref{-1.9}{-1.7}{0.7}{0.0}{new}{\tiny{paper $\ge 2014$}}
    \drawref{-1.9}{-1.9}{0.7}{0.0}{old}{\tiny{paper $<2014$}}
    \end{tikzpicture} 
    \\
    (a)&(b)\\
    \end{tabular}
}
\caption{Labeled network maps, including an example of (1) Vienna metro map with image labels~\cite{wthaly-sedamm-13}, and (b) labeling techniques with respect to their globality and interactiveness.}
\label{fig:tax2}
\end{figure}
\begin{table}[thb]
\vspace*{-3mm}
\caption{The point system for evaluating the criterion effectiveness.}
\centering{
\begin{tabular}{|c|c|l|} \hline
ID & S & Description (x-coordinate in Fig.~\ref{fig:tax2}) \\ \hline

(\textbf{L1})&  & \textbf{Adjustment of metro maps}\\
\lrone{}    & 4     & The layout and label placement is done simultaneously.  \\ \lrtwo{}    & 3     & The layout is uniformly scaled to fit in labels. \\ 
\lrthree{}  & 2     & The directions of the edges are preserved. \\ 
\lrfour{}   & 1     & The layout is not changed when labeling is created. \\ \hline \hline
(\textbf{L2}) & & \textbf{Placement of labels}\\
\lnone{}    & 4     & Uniform placement. \\  
\lntwo{}    & 3     & Criteria for sets of labels. \\ 
\lnthree{}  & 2     & Criteria for pairs labels. \\ 
\lnfour{}   & 1     & Criteria for single labels only. \\ \hline \hline
(\textbf{L2}) & & \textbf{Optimality}\\
\loone{}    & 4     & Global optimality and exact global optimization. \\ 
\lotwo{}    & 3     & Global optimality, but local optimization. \\ 
\lothree{}  & 2     & Local optimality, but global optimality for sub-problems. \\
\lofour{}   & 1     & Local optimality, and local optimization. \\\hline
\end{tabular}
}
\label{tab:score_label}
\end{table}
%
%
In this section we discuss the current state on algorithmic labeling
techniques for metro maps. Similarly to Section~\ref{sec:network}, we
classify the approaches with respect to their degree of globality and
interactivity; see also Figure~\ref{fig:tax2}. We used the following
three criteria to decide on the locality of a labeling approach, including adjustment of metro map \key{(L1)}, placement of labels \key{(L2)}, and optimality \key{(L3)} (see also Table~\ref{tab:score_label}).
For each criterion and each labeling approach we determined the highest score that suits the approach. The overall degree of globality of a labeling approach is then the average over its three assigned scores. 
%
In order to decide on the degree of interactivity for each labeling approach, we applied a similar approach as we did in Section~\ref{sec:network} using Table~\ref{tab:score_time}.
%
We assigned each approach to the lowest category that suits the approach best.  
Altogether, we obtain a measure for both the degree of globality and interactivity of a labeling approach, which we depict in Figure~\ref{fig:tax2}. In the following we discuss the single results in increasing order with respect to their degree of interactivity.    

Nöllenburg and Wolff~\cite{nw-dlhqm-11} present an approach that creates the layout of the metro map and the label placement simultaneously. To that end, they present a mixed integer linear programming formulation that optimally creates a metro map with respect to a global cost-function. They particularly require that labels in between two crossings lie on the same side of the metro line. However, their approach comes with a high running time, e.g., for the metro system of Sydney they report a running time up to $12$ hours. 
Stott et al.~\cite{srmw-amlumo-11} present an iterative approach that creates the layout and label placement in an integrated way using hill climbing based on a global multi-criteria function. Among other criteria they prefer labels on the same side of the metro line. They report running times between 2 and 120 minutes. 

In contrast, Wu et al.~\cite{wthaly-sedamm-13} assume when placing labels that the layout is given. Still, they scale the map to increase free spaces, which they use for label placement based on a MIP approach. By considering the layout and labeling step independently, they achieve running times up to minutes. Yoshida et al.~\cite{yoshida2018} scale single edges of the map while preserving the directions of the edges. The main idea is to create free spaces in dense parts of the map, while leaving other parts of the maps unchanged. 
Similarly to Wu et al.~\cite{wthaly-sedamm-13}, Niedermann and Haunert~\cite{Niedermann2018} systematically scale the map layout to fit in the labeling. For the labeling procedure they propose a dynamic programming approach that labels a single metro line optimally. They use this method as a subroutine of their heuristic, which labels the map within few minutes.

Wu et al.~\cite{wtly-tmla-12} label the stations with photos instead
of text labels. They place the photos alongside two boundaries of the
rectangular map and associate them with their stations via connecting
curves, also called \emph{leaders}. They aim at crossing-free
solutions that minimize the length of the leaders. Wu et
al.~\cite{Wu2015} generalize this approach to all four sides of the
map. Wu et al.~\cite{wtly-zapalmm-11} separate the labels from the
metro map less clearly. They present a genetic algorithm that places
overlap-free text labels and photos within the free spaces of the
metro map connecting the labels with their stations using
straight-line leaders. At the core of the genetic algorithm, which
systematically explores different orders of placing labels, the actual
placement is done by a simple and fast greedy procedure.

Hong et al.~\cite{hmn-avmm-06} separate the placement of the
labels from the computation of the layout. They describe the labeling
problem as a conflict graph based on label candidates and their
occlusions. The label placement corresponds to an independent set of vertices
within the graph, which they determine using simulated
annealing. Although they maximize the number of labels, they cannot
guarantee that each station gets a label.

Wang and Chi~\cite{wc-fmm-11} also separate the placement of labels
from the computation of the metro map layout. They describe the
labeling problem as a multi-criteria energy function, which they
locally optimize.  Their approach particularly avoids occlusions
between labels and prefers labels on the same side of the metro
line. They report running times below one second, which makes their
approach applicable for interactive real-time scenarios. Wang and
Peng~\cite{WangPeng2016} use the same labeling approach for their
interactive editing system.

Chivers and Rodgers~\cite{cr-gidsmd-11} present an approach that after fixing the metro
map layout selects for each station a label out of eight candidates
using a simple greedy approach. They penalize label occlusions and
prefer similarly placed labels of adjacent stations.

Altogether, we observe that with increasing degree of interactivity
the globality of the approaches decreases. In particular, those
approaches that are applicable for interactive real-world scenarios,
typically assume a fixed metro map layout. While this provides the
possibility of computing the label placement much faster than
approaches integrating the process of creating the layout, the given
metro map layout may not host all labels. This is either resolved by
allowing occlusions~\cite{wc-fmm-11}\cite{WangPeng2016} or by labeling only a
(possibly large) subset of stations~\cite{hmn-avmm-06}. Hence, those
approaches are far away from achieving the same quality with respect
to the labeling as approaches with a smaller degree of interactivity. We
therefore deem the development of labeling algorithms with the potential for a high degree of interactivity as a still open and important research problem.

\section{Applications and On-line Tools} \label{sec:app}

Schematic network maps have widely been employed for representing abstract relationships between subjects.
In such cases, 
the schematic maps serve as visual metaphors that effectively systematize the underlying structures hidden behind the target information space.
In this section, 
we first focus on techniques for visualizing abstract information space using schematic map metaphors and then raise representative on-line tools for designing such maps.

\begin{figure}[h]
\centering{

 \setlength{\tabcolsep}{0pt}
 \begin{tabular}{cc}
    \includegraphics[width=0.49\linewidth]{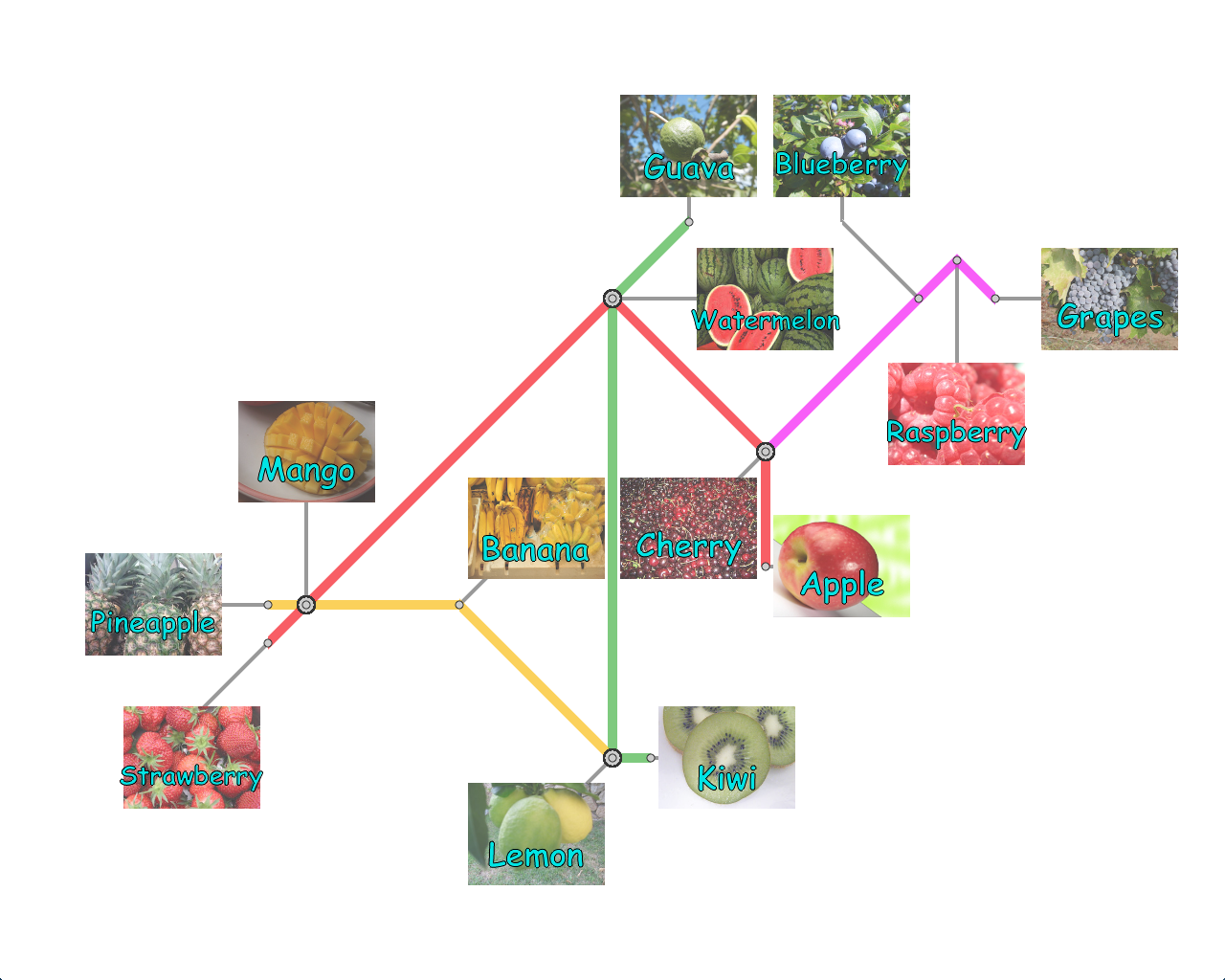} &
    \begin{tikzpicture}[
    scale=1,
    axis/.style={ ->, >=stealth'},
    dashed line/.style={dashed, thin},
    every node/.style={color=black}
    ]
    
    \draw[->] (-1.9,0) node(xline)[below] {\footnotesize Local} -- (1.9,0) node(xline)[below] {\footnotesize Global};
    \draw[->] (0,-1.7) node(yline)[below] {\footnotesize Slow}-- (0,1.7) node(yline)[above] {\footnotesize Fast};
     \drawref{-0.63}{-0.7}{0.0}{-0.2}{old}{\cite{srbms-alppu-2005}} 
     \drawref{-1.48}{-1.5}{0.0}{0.2}{old}{\cite{acp-spvut-2010}}


     \drawref{-1.06}{0.5}{-0.2}{0.2}{old}{\cite{sgh-ttgim-2012}\!\!\cite{sgh-mms-2012}}
     \drawref{-1.06}{1.0}{-0.05}{0.2}{old}{\cite{sysjwl-icczl-2013}}

     \drawref{1.06}{-1.7}{-0.1}{0.15}{new}{\cite{wmbsok-mmpdd-2015}} 
    \drawref{0.63}{-0.7}{0.3}{0.2}{new}{\cite{Oke2015}} 

     \drawref{-1.9}{1.3}{0.0}{-0.18}{old}{\cite{nmpr-tmvcp-2007}}
     \drawref{-1.48}{1.7}{-0.35}{0.1}{old}{\cite{gsz-btcmp-2008}}
     \drawref{-1.9}{-1.7}{0.7}{0.0}{new}{\tiny{paper $\ge 2014$}}
     \drawref{-1.9}{-1.9}{0.7}{0.0}{old}{\tiny{paper $<2014$}}

    \end{tikzpicture}    \\
  (a)&(b)\\
 \end{tabular}
}
\vspace{-3mm}    
\caption{Visualization using schematic maps as metaphors, including (1) an example of fruit maps, and (b) Applications with respect to their globality and interactiveness.}
\label{fig:tax3}
\end{figure}
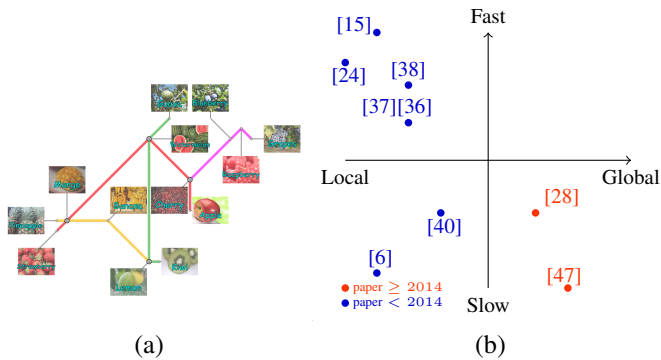

\subsection{Metro Maps as Visual Metaphors}

Effects of using schematic maps as visual metaphors have been investigated, 
though several case studies are at an early stage. 
Nesbitt~\cite{n-gmapu-2004} conducted preliminary studies to evaluate the capabilities of metro map metaphors in understanding abstract data and showed that they can provide a better mental model of the target data as a first impression.
Pontikakis and Twaroch~\cite{pt-smapc-2006} also examined the potential use of metro maps to represent the topological connections between points of interest especially when the underlying topographic maps are unavailable.


Metro map metaphors have often been employed to explore the underlying structure of an information space.
In this case, 
each line corresponds to a chain of subjects such as events, topics, plans, articles, etc.
Stott et al.~\cite{srbms-alppu-2005} visualized the time-lines of project plans and their dependency as a metro map by employing the early version of their hill-climbing algorithm~\cite{srmw-amlumo-11}.
Aguiregoitia et al.~\cite{acp-spvut-2010} applied such an idea to software project visualization where a different set of tasks is aligned with each metro line according to the software development process.

A set of documents is another target for which we can effectively employ a schematic map as a visual metaphor.
Shahaf et al.~\cite{sgh-ttgim-2012} proposed an algorithm for constructing a metro map of documents in such a way that the topology of a document network reflects the desired sparseness inherent in ordinary railway networks.
This technique has been devised to visualize chains of scientific papers to clarify the evolution of research fields~\cite{sgh-mms-2012},
and then further extended to the scalable scheme that enables the the level of detail control in metro map metaphors~\cite{sysjwl-icczl-2013}.

As an application of schematic network metaphors to scientific data,
Wahabzada et al.~\cite{wmbsok-mmpdd-2015} successfully visualized plant disease dynamics by tracking spatiotemporal disease patterns with imaging techniques and composed metro map metaphors by taking advantage of \cite{nw-dlhqm-11}.
Cancer pathways~\cite{hw-smcp-2002} were also visualized as schematic metro maps in the context of multiobjective optimization~\cite{Oke2015}.

Schematic network metaphors also allows us to visualize the structure of low-dimensional space obtained through dimensionality reduction techniques.
Neumayer et al.~\cite{nmpr-tmvcp-2007} employed a metro map metaphor to describe the configuration of segmented characteristic regions over the 2-dimensional plane obtained through self-organizing maps.
Gorban et al.~\cite{gsz-btcmp-2008} applied this idea in the context of dimensionality reduction based on elastic energy minimization to schematize the distribution of data samples with a metro map metaphor.


Several manually designed metaphors of metro maps are available on-line.
Example includes
a Galaxy map that details relationships among planets~\cite{jd-wajts},
a science map composed of scientists and their fields of sciences~\cite{pp-msm},
a human anatomy diagram consisting of organ systems as metro lines~\cite{pch-dchad},
a logistic map of supply chains~\cite{3sm}, 
a map of leading names and domains on the Internet~\cite{ia-wtm4}, 
and a map of digital workspace and marketing technology vendors~\cite{dwmtm}. 
\rv{The most timely example is the Brexit tube maps~\cite{Brexit}, 
which model Brexit issues and sectors as stations and lines, respectively, to elucidate how they relate to one another.}

\subsection{Accessible On-line Tools}

On-line tools often allows us to customize travel routes within the precomputed schematic network maps or even to manually design such schematic maps themselves.
Mapway~\cite{mapway} and ExploreMetro~\cite{exploremetro} are mobile apps for finding a route to the destination on metro maps in major cities.
Metro Map Maker~\cite{mmm} provides a web-based tool for designing schematic network maps,
in which it is possible to manually place stations and draw metro lines by referring to a regular grid on a screen canvas.
Several software libraries for designing interactive schematic maps are also available, including Unfolding Map Library~\cite{uml} for Processing and Java, 
a library for interactive SVG metro maps with JointJS~\cite{cismm},
and a jQuery plug-in for subway map visualization~\cite{smvjp}.


\section{Discussion and Guidance for Followers} \label{sec:discuss}

\rv{
Our hypothesis about the correlation between problem complexity and running time has been visually supported since our scatter plots approximately show a top-left to bottom-right diagonal distribution of algorithmic approaches.
This strongly suggests that researchers are looking for a compromise between the selected constraints for their problem formulation and the techniques to be used, in order to find the balance between the global scope of the solution as opposed to the running time of the algorithm.
Doubtlessly, the final goal of schematic mapping algorithms is to develop an ultimate technique that pinpoints us in the top-right corner of the plots, representing high-quality maps computed instantaneously.
As shown in Figure~\ref{fig:tax1}, the blue markers represent approaches investigated before 2014, and red markers depict the techniques developed after 2014. 
A tendency from left to right and from bottom to top can be observed indicating that researchers are heading towards the desired goal.
While some techniques~\cite{Onda2018}\cite{Oke2015} tend to accelerate the conventional approach by N{\"o}llenburg and Wolff~\cite{nw-dlhqm-11}, which has been considered as the most global optimization method in this paper, the running times are still not very suitable for interactive applications.
This also confirms that the schematic map problem is still a fundamentally difficult one, so that a certain quality loss needs to be tolerated in order to achieve fast performance~\cite{wc-fmm-11}. More sophisticated algorithms are expected to be developed to open further opportunities.
As summarized in Section~\ref{sec:app}, several researchers are beginning to use the maps generated by schematization techniques to visualize their research results, especially when the data is associated with entity connectivity and spatial relationship between entities.
The previous plots can also serve as a visual guidance to instruct those researchers to select an appropriate technique for their research purposes.
}


\section{Future Research Direction} \label{sec:conclude}

In this study, we have compiled the state of the art on schematic network visualization and have given a comprehensive overview of automatic network layout algorithms and map labeling algorithms, as well as tools and applications for use in multidisciplinary domains. 
Our taxonomy demonstrates that most methods carefully design their solutions in compliance with the problem space in order to provide effective approaches with reasonable quality. 
%
Nonetheless until now, no benchmark is provided to enable standard testing on schematic map production, although the Sydney metro map has been studied by most of the conventional map layout algorithms.
Developing such a repository would help promoting the usage of algorithms not only in cartography but also in other scientific domains.
Another challenge for future work in this field is to target advanced methodology to smartly handle global constraints to improve the scalability of the algorithms.
This can be integrated with the interaction techniques together with appropriate user scenarios. 
For this purpose, we plan to further investigate visual representations and interaction schemes in accordance with the task taxonomy for map users.


\section*{Acknowledgment}

The project leading to this submission has received funding from the EU Horizon 2020 research and innovation programme under the Marie Sklodowska-Curie grant No. 747985.

\bibliographystyle{abbrv}
\bibliography{paper}

\begin{thebibliography}{10}

\bibitem{3sm}
3pl subway map.
\newblock
  https://www.supplychainmedia.eu/print/visuals/subway-maps/3pl-subway-map/.

\bibitem{dwmtm}
Digital workplace \& marketing technology vendor map.
\newblock https://www.realstorygroup.com/vendormap/.

\bibitem{exploremetro}
Exploremetro.
\newblock https://www.exploremetro.com/.

\bibitem{mapway}
Mapway.
\newblock https://www.mapway.com/.

\bibitem{ia-wtm4}
Web trend map 4: Coolest gift for geeks.
\newblock https://ia.net/topics/wtm4.

\bibitem{acp-spvut-2010}
A.~Aguirregoitia, J.~J.~D. Cos{\'{i}}n, and C.~Presedo.
\newblock Software project visualization using task oriented metaphors.
\newblock {\em Journal of Software Engineering and Applications},
  3(11):1015--1026, 2010.

\bibitem{cr-gidsmd-11}
D.~Chivers and P.~Rodgers.
\newblock Gesture-based input for drawing schematics on a mobile device.
\newblock In {\em Information Visualisation (IV'11)}, pages 127--134, 2011.

\bibitem{Chivers2014}
D.~Chivers and P.~Rodgers.
\newblock Octilinear force-directed layout with mental map preservation for
  schematic diagrams.
\newblock In T.~Dwyer, H.~Purchase, and A.~Delaney, editors, {\em Diagrammatic
  Representation and Inference (Diagrams'14)}, pages 1--8. Springer Berlin
  Heidelberg, 2014.

\bibitem{Chivers2015}
D.~Chivers and P.~Rodgers.
\newblock Improving search-based schematic layout by parameter manipulation.
\newblock {\em Int. J. Software Engineering and Knowledge Engineering},
  25(6):961--991, 2015.

\bibitem{Brexit}
S.~de~Groot and M.~J. Roberts.
\newblock Brexit mapping.
\newblock \url{http://www.brexitmapping.com}.

\bibitem{jd-wajts}
J.~Diaz.
\newblock We are just a tiny station in the milky way subway map.
\newblock
  https://gizmodo.com/we-are-just-a-tiny-station-in-the-milky-way-subway-map-5454587.

\bibitem{Fink:2014:SMW}
M.~Fink, M.~Lechner, and A.~Wolff.
\newblock Concentric metro maps.
\newblock In {\em Proceedings of the Schematic Mapping Workshop (SMW'14)},
  2014.
\newblock Poster.

\bibitem{fr-gdfp-91}
T.~M.~J. Fruchterman and E.~M. Reingold.
\newblock Graph drawing by force-directed placement.
\newblock {\em Software: Practice and experience}, 21(11):1129--1164, 1991.

\bibitem{g-mbum-94}
K.~Garland.
\newblock {\em Mr. Beck's Underground Map}.
\newblock Capital Transport Publishing, 1994.

\bibitem{gsz-btcmp-2008}
A.~N. Gorban, N.~R. Sumner, and A.~Y. Zinovyev.
\newblock Beyond the concept of manifolds: Principal trees, metro maps, and
  elastic cubic complexes.
\newblock In {\em Principal Manifolds for Data Visualization and Dimension
  Reduction}, pages 219--237. Springer, 2008.

\bibitem{hw-smcp-2002}
W.~C. Hahn and R.~A. Weinburg.
\newblock A subway map of cancer pathways, 2002.
\newblock https://www.nature.com/nrc/poster/subpathways/index.html [Subway map
  designed by C.~Bentley.].

\bibitem{pch-dchad}
B.~P.~C. Ho.
\newblock A doctor created a human anatomy diagram in the style of a subway map
  and it's friggin' gorgeous.
\newblock http://digg.com/2018/human-anatomy-subway-map.

\bibitem{hmn-avmm-06}
S.-H. Hong, D.~Merrick, and H.~A.~D. do~Nascimento.
\newblock Automatic visualisation of metro maps.
\newblock {\em Journal of Visual Languages and Computing}, 17(3):203--224,
  2006.

\bibitem{smvjp}
N.~Kalyani.
\newblock Subway map visualization jquery plugin.
\newblock https://github.com/techbubble/subwayMap.

\bibitem{ld-smagsnm-10}
Z.~Li and W.~Dong.
\newblock A stroke-based method for automated generation of schematic network
  maps.
\newblock {\em Int. J. Geographical Information Science}, 24(11):1631--1647,
  2010.

\bibitem{cismm}
P.~Maszczynski.
\newblock Creating an interactive svg metro map with jointjs.
\newblock
  https://www.netvlies.nl/tips-updates/applicaties/cases/creating-an-interactive-svg-metro-map-with-jointjs/.

\bibitem{uml}
T.~Nagel.
\newblock Unfolding map library.
\newblock http://unfoldingmaps.org/.

\bibitem{n-gmapu-2004}
K.~V. Nesbitt.
\newblock Getting to more abstract places using the metro map metaphor.
\newblock In {\em Information Visualisation}, IV '04, pages 488--493, 2004.

\bibitem{nmpr-tmvcp-2007}
R.~Neumayer, R.~Mayer, G.~Polzlbauer, and A.~Rauber.
\newblock The metro visualisation of component planes for self-organising maps.
\newblock In {\em Int. Joint Conference on Neural Networks}, pages 2788--2793,
  2007.

\bibitem{Niedermann2018}
B.~Niedermann and J.~Haunert.
\newblock An algorithmic framework for labeling network maps.
\newblock {\em Algorithmica}, 80(5):1493--1533, 2018.

\bibitem{noellenburg:2014:schematicmapping}
M.~N{\"o}llenburg.
\newblock A survey on automated metro map layout methods.
\newblock In {\em Schematic Mapping Workshop}, Essex, UK, 2014.
\newblock \url{https://i11www.iti.kit.edu/extra/publications/n-asamm-14.pdf}.

\bibitem{nw-dlhqm-11}
M.~N{\"o}llenburg and A.~Wolff.
\newblock Drawing and labeling high-quality metro maps by mixed-integer
  programming.
\newblock {\em {IEEE} Transactions Visualization and Computer Graphics},
  17(5):626--641, 2011.

\bibitem{Oke2015}
O.~Oke and S.~Siddiqui.
\newblock Efficient automated schematic map drawing using multiobjective mixed
  integer programming.
\newblock {\em Computers \& Operations Research}, 61:1--17, 2015.

\bibitem{Onda2018}
M.~Onda, M.~Moriguchi, and K.~Imai.
\newblock Automatic drawing for tokyo metro map.
\newblock In {\em European Workshop on Computational Geometry (EuroCG'18)},
  pages 62:1--62:6, 2018.

\bibitem{o-mmw-03}
M.~Ovenden.
\newblock {\em Metro Maps of the World}.
\newblock Capital Transport Publishing, 2003.

\bibitem{pp-msm}
P.~Plait.
\newblock Metrocontextual science map.
\newblock
  \url{http://blogs.discovermagazine.com/badastronomy/2010/08/31/metrocontextual-science-map/\#.XFEYW88zbfB}.

\bibitem{pt-smapc-2006}
E.~Pontikakis and F.~Twaroch.
\newblock Schematic maps as an alternative to point coverages when topographic
  maps are not available.
\newblock In {\em Information Visualisation}, IV '06, pages 297--303, 2006.

\bibitem{r-umu-12}
M.~J. Roberts.
\newblock {\em Underground Maps Unravelled: Explorations in Information
  Design}.
\newblock Self-published, 2012.

\bibitem{Roberts:2017:HCS}
M.~J. Roberts, H.~Gray, and J.~Lesnik.
\newblock Preference versus performance: Investigating the dissociation between
  objective measures and subjective ratings of usability for schematic metro
  maps and intuitive theories of design.
\newblock {\em International Journal of Human-Computer Studies}, 98:109 -- 128,
  2017.

\bibitem{Roberts:2013:HCS}
M.~J. Roberts, E.~J. Newton, F.~D. Lagattolla, S.~Hughes, and M.~C. Hasler.
\newblock Objective versus subjective measures of paris metro map usability:
  Investigating traditional octolinear versus all-curves schematics.
\newblock {\em International Journal of Human-Computer Studies}, 71(3):363 --
  386, 2013.

\bibitem{sgh-mms-2012}
D.~Shahaf, C.~Guestrin, and E.~Horvitz.
\newblock Metro maps of science.
\newblock In {\em Knowledge Discovery and Data Mining}, KDD '12, pages
  1122--1130, 2012.

\bibitem{sgh-ttgim-2012}
D.~Shahaf, C.~Guestrin, and E.~Horvitz.
\newblock Trains of thought: Generating information maps.
\newblock In {\em World Wide Web}, WWW '12, pages 899--908, 2012.

\bibitem{sysjwl-icczl-2013}
D.~Shahaf, J.~Yang, C.~Suen, J.~Jacobs, H.~Wang, and J.~Leskovec.
\newblock Information cartography: Creating zoomable, large-scale maps of
  information.
\newblock In {\em Knowledge Discovery and Data Mining}, KDD '13, pages
  1097--1105, 2013.

\bibitem{srmw-amlumo-11}
J.~Stott, P.~Rodgers, J.~C. Martínez-Ovando, and S.~G. Walker.
\newblock Automatic metro map layout using multicriteria optimization.
\newblock 17(1):101--114, 2011.

\bibitem{srbms-alppu-2005}
J.~M. Stott, P.~Rodgers, R.~A. Burkhard, M.~Meier, and M.~T.~J. Smis.
\newblock Automatic layout of project plans using a metro map metaphor.
\newblock In {\em Information Visualisation}, IV '05, pages 203--206, 2005.

\bibitem{sm-gdmsm-95}
K.~Sugiyama and K.~Misue.
\newblock Graph drawing by the magnetic spring model.
\newblock {\em Journal of Visual Languages \& Computing}, 6(3):217--231, 1995.

\bibitem{Ti2014}
P.~Ti and Z.~Li.
\newblock Generation of schematic network maps with automated detection and
  enlargement of congested areas.
\newblock {\em International Journal of Geographical Information Science},
  28(3):521--540, 2014.

\bibitem{Ti2015}
P.~Ti, Z.~Li, and Z.~Xu.
\newblock Automated generation of schematic network maps adaptive to display
  sizes.
\newblock {\em The Cartographic Journal}, 52(2):168--176, 2015.

\bibitem{mmm}
S.~Turner.
\newblock Metro map maker.
\newblock https://metromapmaker.com/.

\bibitem{Dijk2018}
T.~C. van Dijk and D.~Lutz.
\newblock Realtime linear cartograms and metro maps.
\newblock In {\em Advances in Geographic Information Systems (SIGSPATIAL'18)},
  pages 488--491. ACM, 2018.

\bibitem{Dijk2014}
T.~C. van Dijk, A.~van Goethem, J.-H. Haunert, W.~Meulemans, and B.~Speckmann.
\newblock Map schematization with circular arcs.
\newblock In M.~Duckham, E.~Pebesma, K.~Stewart, and A.~U. Frank, editors, {\em
  Geographic Information Science (GIScience'14)}, pages 1--17. Springer
  International Publishing, 2014.

\bibitem{wmbsok-mmpdd-2015}
M.~Wahabzada, A.-K. Mahlein, C.~Bauckhage, U.~Steiner, E.-C. Oerke, and
  K.~Kersting.
\newblock Metro maps of plant disease dynamics-automated mining of differences
  using hyperspectral images.
\newblock {\em PLOS ONE}, 10(1):1--20, 2015.

\bibitem{wc-fmm-11}
Y.-S. Wang and M.-T. Chi.
\newblock Focus+context metro maps.
\newblock {\em {IEEE} Transactions Visualization and Computer Graphics},
  17(12):2528--2535, 2011.

\bibitem{WangPeng2016}
Y.-S. Wang and W.-Y. Peng.
\newblock Interactive metro map editing.
\newblock {\em IEEE Transactions on Visualization and Computer Graphics},
  22(2):1115--1126, 2016.

\bibitem{w-dsms-07}
A.~Wolff.
\newblock Drawing subway maps: A survey.
\newblock {\em Informatik -- Forschung und Entwicklung}, 22(1):23--44, 2007.

\bibitem{w-gdc-13}
A.~Wolff.
\newblock Graph drawing and cartography.
\newblock In R.~Tamassia, editor, {\em Handbook of Graph Drawing and
  Visualization}, chapter~23, pages 697--736. CRC Press, 2013.

\bibitem{Wu2015}
H.-Y. Wu, S.-H. Poon, S.~Takahashi, M.~Arikawa, C.-C. Lin, and H.-C. Yen.
\newblock Designing and annotating metro maps with loop lines.
\newblock In {\em Information Visualization}, IV '15, pages 9--14. IEEE, 2015.

\bibitem{wthaly-sedamm-13}
H.-Y. Wu, S.~Takahashi, D.~Hirono, M.~Arikawa, C.-C. Lin, and H.-C. Yen.
\newblock Spatially efficient design of annotated metro maps.
\newblock {\em Computer Graphics Forum}, 32(3):261--270, 2013.

\bibitem{wtly-zapalmm-11}
H.-Y. Wu, S.~Takahashi, C.-C. Lin, and H.-C. Yen.
\newblock A zone-based approach for placing annotation labels on metro maps.
\newblock In {\em Smart Graphics (SG'11)}, volume 6815 of {\em LNCS}, pages
  91--102. Springer-Verlag, 2011.

\bibitem{wtly-tmla-12}
H.-Y. Wu, S.~Takahashi, C.-C. Lin, and H.-C. Yen.
\newblock Travel-route-centered metro map layout and annotation.
\newblock {\em Computer Graphics Forum}, 31(3):925--934, 2012.

\bibitem{yoshida2018}
Y.~Yoshida, K.~Maruyama, T.~Kawagoe, H.-Y. Wu, M.~Arikawa, and S.~Takahashi.
\newblock Progressive annotation of schematic railway maps.
\newblock In {\em Information Visualisation}, IV '18, pages 373--378. IEEE,
  2018.

\end{thebibliography}

\end{document}